# A Real-Time Remote-Sensing-Guided Decision-Support Framework for Cloud-Seeding Operations: A Field Demonstration Using Himawari-9 and C-band Phased Array Weather Radar

Shunji Kotsuki[1,2,3]

Corresponding author

Email: shunji.kotsuki@chiba-u.jp

ORCID id: 0000-0002-0717-9192

Kazuaki Yasunaga[4]

Email: yasunaga@sus.u-toyama.ac.jp

ORCID id: 0000-0003-2263-4939

Atsushi Hamada[4]

Email: hamada@sus.u-toyama.ac.jp

ORCID id: 0000-0002-7031-0245

Kazuhiro Yoshimi[5]

Email: kyoshimi@pu-toyama.ac.jp

ORCID id: 0009-0000-1156-2795

Kenta Shiraishi[6]

Email: kenta.shiraishi@chiba-u.jp

ORCID id: 0009-0006-5769-8122

Akira Takeshima[1]

Email: takeshima@chiba-u.jp

ORCID id: 0009-0006-5769-8122

Yoshiki Terashima[7]

Email: y.terashima@ana.co.jp

ORCID id: N/A

Kaya Kanemaru[8]

Email: kanemaru@nict.go.jp

ORCID id: 0000-0003-0352-323X

Yusuke Hiraga[9]

Email: yusuke.hiraga.c3@tohoku.ac.jp

ORCID id: 0000-0002-7791-5431

Takuya Funatomi[10]

Email: funatomi.takuya.2c@kyoto-u.ac.jp

ORCID: 0000-0001-5588-5932

Hiroyuki Kubo[11]

Email: hkubo@chiba-u.jp

ORCID: 0000-0002-7061-7941

Kohei Suzuki[12]

Email: kohei@tgys.co.jp

ORCID: N/A

Masanao Ohashi[1]

Email: ohash1906@chiba-u.jp

ORCID id: 0000-0002-8553-5308

Tadashi Tsuyuki[1]

Email: ttsuyuki@chiba-u.jp

ORCID id: 0009-0009-5365-6226

(Institutional addresses)

[1] Center for Environmental Remote Sensing, Chiba University, Yayoi-Cho 1-33, Inage-Ku, Chiba 263-8522, Japan

[2] Institute for Advanced Academic Research, Chiba University, Yayoi-Cho 1-33, Inage-Ku, Chiba 263-8522, Japan

[3] Research Institute of Disaster Medicine, Chiba University, Yayoi-Cho 1-33, Inage-Ku, Chiba 263-8522, Japan

[4] Department of Earth Science, Graduate School of Science and Engineering, University of Toyama, Toyama, 930-8555, Japan

[5] Department of Environmental and Civil Engineering, Toyama Prefectural University, 5180

Kurokawa, Imizu, Toyama, 939-0398, Japan

[6] Graduate School of Science and Engineering, Chiba University, Yayoi-Cho 1-33, Inage-Ku, Chiba, 263-8522, Japan

[7] ANA Holdings Inc., Shiodome City Center, 1-5-2 Higashi-Shimbashi, Minato-ku, Tokyo 105-7140, Japan

[8] Radio Research Institute, National Institute of Information and Communications Technology, 4-2-1 Nukuikita-machi, Koganei, Tokyo, Japan

[9] Department of Civil and Environmental Engineering, Tohoku University, Aoba-yama 6-6-06, Aoba, Sendai, 980-8579, Japan

[10] Academic Center for Computing and Media Studies, Kyoto University, 606-8501, Japan

[11] Graduate School of Informatics, Chiba University, Yayoi-Cho 1-33, Inage-Ku, Chiba, 263-8522, Japan

[12] Tagayasu Inc., 181 Koizumi-Cho, Toyama-Shi Toyama, 939-8082, Japan

## Abstract

Cloud seeding has recently attracted renewed attention as a potential component of future weather-intervention strategies for mitigating high-impact hydrometeorological events. However, aircraft-based cloud-seeding operations targeting rapidly evolving clouds require real-time identification of suitable target clouds and operational guidance under practical constraints, such as observational latency, limited aircraft visibility, and restricted communication.

This study proposes a real-time remote-sensing-guided decision-support framework for cloud-seeding operations using high frequency geostationary satellite and ground weather radar observations. The framework integrates cloud assessment, human-in-the-loop decision support, and aircraft operation to translate high-frequency remote-sensing information into actionable guidance for seeding aircraft. We demonstrate the framework using 2.5-min Himawari-9 geostationary satellite observations and 60-s C-band phased-array weather radar (C-PAWR) observations during the preliminary dry-ice cloud-seeding field campaign conducted over Toyama Bay, Japan, in January 2026. In the 13 January case, the framework enabled the ground team to identify a developing cumulus cloud with a lifetime of approximately 20 min, communicate guidance to the aircraft, and conduct seeding immediately before the cloud began to dissipate naturally. Candidate seedable clouds were identified from Himawari-9 infrared indices, and their selection was supported by near-real-time C-PAWR observations of precipitation echoes. Because the released dry-ice amount was limited to 30 kg, this study does not attempt to attribute subsequent cloud evolution to seeding effects. Instead, the results demonstrate that rapid-scan satellite and ground radar observations can support real-time target selection and aircraft guidance for responsible, operationally feasible weather-intervention field experiments.

# 1 Introduction

Climate change is increasing the urgency of strengthening resilience to weather-related disasters, including torrential rainfall, tropical cyclones, and other high-impact hydrometeorological extremes (Papalexiou and Montanari, 2019; IPCC 2021; Seneviratne et al. 2021). Existing approaches to climate adaptation and disaster risk reduction have primarily focused on improving forecast accuracy, early-warning systems, and structural and non-structural protection countermeasures (Kreibich et al. 2015). While these countermeasures remain indispensable, the increasing complexity of weather-related risks motivates the exploration of complementary methods and tools that may enhance the resilience of social systems to climate change.

Weather modification has been explored as a complementary approach for mitigating high-impact weather phenomena, particularly tropical cyclones and hurricanes (Alamaro et al., 2006; Klima et al., 2012). Historical examples include Project Cirrus (1947–1952) and Project Stormfury (1962–1983), which attempted to weaken hurricanes through cloud seeding by introducing ice-nucleating particles into convective cloud systems (Willoughby et al., 1985; Abe et al., 2025). However, the scientific validation of such effects proved difficult because of the large natural variability of atmospheric systems, the limited ability to observe the relevant physical processes, and the challenge of attributing observed storm changes to the seeding intervention itself, leading to discontinuation of the project (Willoughby et al., 1985). However, recent advances in observational technologies, data assimilation, and numerical modeling have motivated renewed theoretical and practical interest in the possibility of mitigating extreme weather through targeted weather modification or control (Huang et al. 2025; Miyoshi 2026).

Recently, Japan's Moonshot Research and Development Program has therefore begun exploring weather intervention as a potential long-term option for mitigating extreme weather impacts (Boyd 2023). The program aims to develop technologies for guiding weather systems toward less damaging states through interventions (or modifications), including the optimization of intervention strategies for controlling chaotic dynamical systems (e.g. Miyoshi and Sun 2022; Kawasaki and Kotsuki 2024; Sawada 2024; Kurosawa et al. 2025). In this Moonshot Program, cloud seeding is positioned as a major weather modification approach that would be effective for mitigating stationary and extreme heavy rainfall.

Cloud seeding has been widely studied as a weather modification technique mainly for augmenting precipitation and securing water resources (Bruintjes, 1999; Flossmann et al., 2019). Operational weather modification programs, including fog dispersion, rain and snow enhancement, and hail suppression, have been conducted in many countries worldwide (WMO, 2025). Cloud seeding is broadly classified into glaciogenic and hygroscopic seeding, with glaciogenic seeding having been reported to enhance precipitation, particularly in orographic cloud systems (French et al., 2018; Tessendorf et al., 2019; Flossmann et al., 2019). Recent field programs have also emphasized the importance of combining seeding operations with coordinated aircraft, radar, and in situ observations to evaluate both physical responses and operational feasibility (e.g. Tessendorf et al., 2019; Prabhakaran et al. 2023). More recently, numerical studies have suggested that overseeding, in which a high concentration of ice-nucleating particles is intentionally introduced, may modify cloud microphysical processes and potentially mitigate intense rainfall (Hiraga et al. 2026).

Successful cloud seeding for rapidly evolving intense rainfall, however, requires not only an understanding of the underlying cloud microphysical processes but also operational capabilities to identify and target individual clouds using observations and numerical simulations. Although previous studies have substantially advanced the physical understanding and evaluation of cloud-seeding effects (French et al., 2018; Tessendorf et al., 2019; Flossmann et al., 2019) and seedability assessments (Hashimoto et al. 2008), comparatively less explicit attention has been paid to the real-time decision-support process required to identify suitable clouds and guide seeding aircraft under practical field constraints. This gap is particularly important in Japan's Moonshot Research and Development Program, which explores the possibility of conducting cloud seeding over the ocean before water vapor is transported into land areas, thereby requiring operational knowledge for targeting specific cloud systems using aerial platforms such as aircraft or drones.

In this study, we propose a real-time remote-sensing-guided decision-support framework for cloud-seeding operations, based on high-frequency observations from the geostationary satellite and ground weather radar observations. Here we used Himawari-9 and a C-band phased-array weather radar (C-PAWR) observations, and explored conducting cloud-seeding operations under multiple real-world constraints. For example, although Himawari-9 observations are available every 2.5 min over Japan, the corresponding imagery reaches the operation team with a delay of approximately 5–10 min. In addition, the clouds visible from the aircraft are limited, and communication between the aircraft and the ground teams was restricted to voice calls. The objective of this study is therefore twofold: to propose an operational framework for real-time cloud-seeding support, and to clarify the practical requirements, bottlenecks, and field campaign-based lessons for conducting responsible and operationally feasible weather-intervention field experiments. The present study reports operational lessons learned from a preliminary field experiment conducted over Toyama Bay in January 2026 as part of Japan’s Moonshot Research and Development Program. Note that this study does not aim to demonstrate a statistically significant meteorological modification effect. Instead, it aims to clarify the practical requirements and bottlenecks associated with aircraft-based cloud-seeding operations under realistic field constraints, and to demonstrate how high-frequency remote sensing data can support real-time target selection and operational decision making. In addition to these technical contributions, this paper also reports the procedural steps taken to conduct a cloud-seeding experiment in Japan, because planning field experiments involving intentional weather modification requires careful consideration of ethical, legal, and social issues (ELSI). Sharing such procedures provides important practical knowledge for future responsible weather-intervention research in other countries as well.

The rest of the paper is organized as follows. Section 2 provides an overview of the field campaign. Section 3 presents the proposed remote-sensing-based decision-support framework. Section 4 demonstrates its application to the cloud-seeding operations and aircraft observations conducted during the campaign. Finally, Section 5 provides a summary.

## 2 Field Campaign and Operational Requirements

### 2.1 Campaign Setting

The field campaign was conducted in early January 2026 over Toyama Bay (hereafter Toyama Bay 2026 campaign). Toyama Bay was selected because wintertime snow clouds associated with Sea-of-Japan winter monsoon systems frequently develop in this region, and previous aircraft and in situ observations have shown that such snow clouds can contain substantial supercooled liquid water under favorable conditions (Murakami et al., 1994;

Kusunoki et al., 2005; Murakami, 2019). In addition, the area is covered by the C-PAWR observations (Figure 1 a), and the downstream advection of possible seeding-induced effects was expected to remain primarily over the ocean, reducing potential safety concerns over land.

The Toyama Bay 2026 campaign was implemented through a coordinated operation between an aircraft team and a ground team. Such coordination between aircraft operations and ground-based observational support is increasingly recognized as important in modern cloud-seeding field programs, as demonstrated in previous aircraft- and radar-based experiments such as SNOWIE and CAIPEEX (Tessendorf et al. 2019; Prabhakaran et al. 2023). The ground team served as the decision-support unit, responsible for integrating ground-based observations and remote-sensing data to assess operational seedability and identify candidate target clouds. A ground observation site was established near the planned seeding area to monitor local meteorological conditions (Figure 1 b). The resulting seedability information was communicated to the aircraft team, which used it to adjust the flight route and conduct dry-ice seeding near the selected target clouds.

## 2.2 Ground-based and remote-sensing observations

The field campaign was supported by both ground-based observations and high-frequency remote-sensing observations. For remote-sensing-based operational support, we mainly used every 2.5-min Himawari-9 geostationary satellite and every 60-s C-PAWR observations. Himawari-9 satellite provides broad-area information on cloud distribution and cloud-top conditions, whereas C-PAWR provides high-frequency, three-dimensional information on the evolution of clouds and precipitation systems around the target area. These complementary observations formed the basis for real-time seedability assessments and target-cloud selection. The high-frequency 3D C-PAWR observations were also used for safety monitoring, allowing the ground team to track precipitation evolution around the target area and check for any unexpected changes following the seeding operations.

In addition to the available remote-sensing data, we set a ground observation site for Go/No-Go decisions, operational decisions on whether to conduct or cancel the seeding experiment based on meteorological, safety, and logistical conditions (see section 2.4 for more details). At the ground observation site, meteorological and cloud-related conditions were monitored using an automatic weather station (known as AWS), a vertically pointing K-band radar, a lidar system, and live cameras (Figure 1 b). These instruments provided local information on near-surface meteorological conditions, vertical cloud and precipitation structures, and visual cloud evolution around the experimental area. The ground observational site was located at Sonokeyama Camp at Nyuzen, Toyama (36.93ºN, 137.44ºE).

Among the ground-based instruments, the lidar and K-band radar played important roles in the campaign. These observations were used to monitor the vertical structure of clouds and precipitation, especially to estimate cloud-top height around the experimental area. In the present campaign, the aircraft altitude available for seeding was limited to approximately the 700-hPa level, because the cabin was exposed to outside air pressure during the seeding operations. Therefore, the lidar and K-band radar observations were used to assess whether the target clouds were located below the operational altitude limit of the aircraft.

### 2.3 Aircraft-based seeding operation

Cloud seeding was conducted using a King Air 200T aircraft operated by Diamond Air Service Inc. The King Air 200T is a twin-turboprop aircraft that has been used for research and observational missions, including atmospheric and meteorological measurements in Japan. The aircraft was equipped with a ventilation port that can be used for meteorological observations and sampling. This campaign employed dry-ice seeding, which is a classical glaciogenic seeding approach that promotes ice formation in supercooled clouds through rapid cooling and ice nucleation (Schaefer, 1946; Fukuta et al., 1971). In this campaign, dry-ice pellets were released from inside the aircraft through this port.

The seeding operation was conducted under two major constraints associated with aircraft operation in Japan. First, internet communication was not available onboard the aircraft. Second, communication between the aircraft and ground teams was limited to voice calls via Iridium. Therefore, real-time seedability information was analyzed by the ground team and transmitted verbally to the aircraft team. The aircraft team then used this information, together with visual observations from the aircraft, to adjust the flight operation and conduct the seeding.

### 2.4 Stakeholder Communication and Operational Governance

As described in Section 2.1, field experiments involving weather modification have been rare in Japan since the Japanese Cloud Seeding Experiments for Precipitation Augmentation (JCSEPA), a national research project led by the Meteorological Research Institute and collaborating institutions during 2006–2011 (Murakami and JCSEPA Research Group, 2011; Kuo et al., 2024). Therefore, the Toyama Bay 2026 campaign was implemented with careful attention to ELSI (or responsible research and innovation; RRI; e.g. Stilgoe et al. 2020). In particular, the project team conducted sufficient prior communication with local stakeholders. Before the experiment, coordination meetings and briefings were held with local governments and relevant public agencies (local meteorological authorities and the Japan Air Self-Defense Force), and information on the planned experiment was shared with related organizations. Two public briefing sessions were also organized for local residents and other stakeholders. In addition, two press releases were issued before the start of the campaign to inform the wider public through the media. From a scientific and safety perspective, the field campaign was reviewed through the experimental safety review process established within the Moonshot Research and Development Program as well. In addition, prior to the field campaign, numerical experiments were conducted using the NWP models to assess the potential impact of the planned dry-ice seeding using two cloud microphysical schemes involving cloud seeding. Specifically, the first used The Weather Research and Forecasting (WRF) model (Powers et al. 2017) with a two-moment bulk microphysics scheme that represents cloud-seeding processes (Hiraga et al., 2026). The second used the Advanced Microphysics Prediction System (Hashino et al. 2020) which is a bin microphysics scheme incorporating seeding parameterization (Hashimoto and Murakami 2016). These simulations indicated that the release of 30 kg of dry ice was unlikely to produce a substantial impact on the snowfall system. The field experiment was therefore conducted after confirming that the planned seeding scale was sufficiently limited from a safety and environmental perspective.

Operational governance was also emphasized in the Go/No-Go decision process. Before the campaign, explicit criteria for conducting or canceling the experiment were defined and made publicly available on the project website to enhance transparency. The final decision to conduct seeding was made through a briefing process held on the day before the operation and again on the early morning of the day of operations. Whether the experiment was conducted or canceled, the decision was communicated by email to relevant organizations. When seeding operations were conducted, a brief report was posted on the project website on the same day, and information was also shared with the Cabinet Office, the Ministry of Education, Culture,

Sports, Science and Technology, and the Japan Science and Technology Agency, which oversee the Moonshot Program. These procedures were implemented to ensure safety, improve procedural transparency, and address public concerns associated with intentional weather modification experiments.

# 3 Decision Support Framework

## 3.1 Overview of the architecture

In this study, we propose a real-time remote-sensing-guided decision-support framework for aircraft-based cloud-seeding operations (Figure 2). The framework consists of three main components: cloud assessment, decision support, and seeding operation. Its purpose is not only to visualize high-frequency observations, but also to translate observation data into actionable guidance for aircraft-based seeding under realistic operational constraints.

Before conducting seeding experiments, the aircraft and ground teams first hold joint briefings for Go/No-Go decisions based on the operational criteria described in Section 2.4. The aircraft operators also participated in these briefings and contributed to the Go/No-Go decisions from the perspective of flight safety, including visual flight conditions and lightning risk. Once the seeding experiment was decided, operational seedability was assessed in real time using Himawari-9 and C-PAWR observations, as described in Sections 3.2 and 3.3. Based on this seedability assessment, the ground team provided seedability guidance to the aircraft team as part of the decision-support process (Section 3.4). The aircraft team then conducted the seeding operation while also providing operational feedback to the ground team. This feedback loop allowed the ground and aircraft teams to continuously update their situational awareness and improve the cloud assessment system and upcoming seeding operations.

### 3.2 Seedability assessment from Himawari-9 Geostationary Satellite

Operational seedability was assessed using geostationary satellite Himawari-9 observations, which is a Japanese geostationary meteorological satellite that continuously observes the Asia-Pacific region (Bessho et al. 2016). Over the Japan region, Himawari-9 satellite provides high-frequency observations every 2.5 min and high spatial resolution (0.5 km for the visible red band, 1.0 km for visible green and blue bands, and 2 km for other three near infrared and ten infrared bands). This high spatial and temporal resolution of Himawari-9 observations enabled the ground team to monitor the development, movement, and dissipation of target clouds over Toyama Bay.

In this study, Himawari-9 observations played the central role in identifying candidate clouds for seeding and to support real-time operational seedability assessment. The present campaign focused on glaciogenic seeding, for which identifying clouds containing supercooled liquid water is essential. Wintertime clouds with supercooled liquid water are known to occur over Toyama Bay, particularly within cumulus clouds associated with upward plumes. In situ observations during the campaign, including HYVIS (HYdrometeor VideoSonde) measurements, confirmed the presence of clouds containing abundant supercooled liquid water (Dr. Kanada, personal communications). Because operational seeding requires such clouds to be detected before clouds' dissipations, Himawari-9 observations play a central role in the seedability assessment in this field campaign. While Himawari-9 satellite does not directly provide the vertical distribution of supercooled liquid water, its 2.5-min imagery over Japan enables near-real-time monitoring of the formation, growth, and movement of candidate cumulus clouds before clear radar precipitation signatures appear.

For the Himawari-9-based seedability assessment, we used three indices calculated from two infrared brightness temperatures: Band 13 (hereafter B13; 10.4073 μm) and Band 15 (hereafter B15; 12.3806 μm). These two infrared window channels have different absorption characteristics associated with water vapor and cloud particles. The first index is B13 which provides information on cloud-top brightness temperature, and therefore a proxy for cloud-top height. The second index is temporal difference in B13 (hereafter Tdiff in B13) in which negative (positive) temporal difference indicates cloud growth (dissipation). The third index is difference between B13 and B15. For optically thin (thick) clouds, the brightness temperatures observed in the two channels tend to result in a positive (near zero) B13−B15 (Inoue 1987). Because the target clouds for glaciogenic seeding in this campaign were expected to be optically thick cumulus clouds, small B13−B15 values (< 1.0 K) were used as an indicator of candidate seedable clouds. See Appendix A for further details on the web-based seedability assessment tool.

Here, we demonstrate how the seedability assessment system can support target interpretation by comparing aircraft visual observations with Himawari-9-based visible imagery. Figure 3 shows an example from the 13 January seeding operation, which is discussed in detail in Section 4. For comparison with the assessment system, we use camera images taken after the seeding operation, when the aircraft had climbed to 15,000 ft (4,572 m), together with the corresponding Himawari-9 visible image. The numbered cloud features in the aircraft images correspond to cloud elements identified in the Himawari-9 visible image. Figure 4 shows the corresponding visualization results from the seedability assessment system. Comparing Figures 3 and 4 provides further insight into the characteristics and utility of the seedability assessment system.

In Figure 3 (b-1), the cold front is clearly captured in the Himawari-9 visible image. Subsequent visual observations from the aircraft suggested that the cloud-top height associated with the cold front was approximately 20,000 ft (6,096 m). Behind the cold front, cloud features with distinct shadows can be observed, whereas diffuse cloudiness spreads ahead of the cold front. This diffuse cloud region is interpreted as an anvil cloud associated with frontal convection. The anvil advected ahead of the surface cold front by strong upper-level westerly winds, causing it to appear on the forward side of the frontal cloud band. This anvil cloud can also be identified in the aircraft visual observations shown in Figure 3 (a). In Figure 4 (a), the high-altitude convective cloud associated with the cold front is characterized by low B13, reflecting cold cloud-top temperatures. The small brightness temperature difference between B13−B15 suggests that the cloud was optically thick. Furthermore, negative Tdiff in B13 extends ahead of the cold front, suggesting cloud-top cooling associated with the eastward propagation and ongoing development of the frontal cloud system.

The clouds targeted in the present experiment were low-level clouds with cloud-top heights below approximately 3,000 m, such as the cloud features numbered 1–7 in Figure 3. Although these clouds could be visually identified from the aircraft, it was difficult to evaluate their detailed characteristics by visual inspection alone. By contrast, the seedability assessment system was able to highlight relatively higher cloud elements, such as features 1, 2, and 6 (Figure 4 a). Although these clouds did not satisfy the optical-thickness criterion (B13-B15 < 1.0 K) in this particular example, we later demonstrate that developing target clouds can satisfy the second and third seedability criteria (Figure 7).

It should be noted that visual seedability assessment from the aircraft becomes more difficult at the actual seeding altitude. The cloud system indicated by feature 1 in Figure 3 was the target cloud system seeded in this experiment, and its cloud-top height was approximately 7,000 ft (2,134 m). When the aircraft descended to the seeding altitude of cloud top, it was no longer possible to observe the cloud field from a broad, bird's-eye perspective. In practice, therefore, seedability reports provided by the ground team based on remote-sensing observations were essential for conducting effective seeding operations.

### 3.3 Assessments with 60-s C-band phased-array weather radar

In addition to the broad-area cloud-top information provided by Himawari-9 data, this study used the C-PAWR installed by Japan Radio Co., Ltd. (JRC) in Namerikawa, Toyama. Here, PAWR observations provide dense and frequent volumetric information that is particularly useful for monitoring rapidly evolving convective precipitation systems (Otsuka et al., 2016; Palmer et al., 2022). The C-PAWR used in this campaign is an experimental system under development and is a small-scale model of a C-band operational radar. In terms of observational specifications, the radar operates in the 5-GHz C-band, with a range resolution of approximately 125 m and an observation range of approximately 60 km.

In this study, the final target selection was made by the ground team by integrating seedability information from Himawari-9 and C-PAWR observations. During the field campaign, C-PAWR data were available to the ground team with almost no operational latency, enabling near-real-time confirmation of precipitation echoes around candidate clouds. Here, every 60-s C-PAWR observations were used as supporting information to evaluate whether the candidate clouds were accompanied by precipitation systems.

### 3.4 Human-in-the-loop target selection and decision support

The decision-support framework proposed in this study was designed not merely for the automatic extraction of candidate clouds, but as a human-in-the-loop decision-support system that enables the ground team to integrate multiple sources of observational information and translate them into actionable guidance for the aircraft team. Because clouds targeted for seeding evolve rapidly in both space and time, it is difficult to determine target clouds based on a single source of observational information.

During the operation, the ground team selected candidate clouds while mainly monitoring the seeding assessment system based on Himawari-9 observations. The selection process considered not only whether the clouds exhibited characteristics suitable for seeding, but also whether the aircraft could reach the target within the available time, and whether the candidate clouds were located within the authorized seeding airspace (the cyan polygon of Figure A1).

During the experiment, internet communication was not available onboard the aircraft, and communication between the aircraft and ground teams was limited to voice calls. Therefore, the seedability information analyzed by the ground team had to be translated into concise instructions that could be immediately understood by the aircraft team. Information such as the location and movement direction of the target cloud, the target area, and the timing of seeding was communicated by voice. The aircraft team then adjusted the flight route and conducted seeding of the target cloud by combining the guidance from the ground team with visual observations from the aircraft. In this way, the proposed framework connected high-frequency remote-sensing information to executable aircraft operations through interpretation and decision making by the ground team. This enabled the framework to support target-cloud selection and aircraft guidance even under practical operational constraints, including observational data latency, the limited range of clouds visible from the aircraft, and communication limitations.

# 4 Field Demonstration

## 4.1 Overview of operations

Table 1 summarizes the daily operations conducted during the Toyama Bay 2026 campaign. The campaign period extended from 6 to 14 January 2026, during which seeding operations were conducted on four days: 7, 10, 12, and 13 January. On the remaining days, operations were canceled based on the Go/No-Go criteria described in Section 2.4, including wind direction, weather conditions, and lightning risk. The conducted operations covered different cloud conditions, including cloud-free-region seeding, stationary-cloud seeding, stratiform-cloud seeding, and convective-cloud seeding.

Among these cases, the 13 January operation provided the clearest example of real-time target selection and aircraft guidance based on the proposed framework, which is therefore examined in detail below. This case is particularly relevant to the objective of this study because, as discussed in the Introduction, the proposed framework aims to support real-time cloud seeding of developing precipitation systems, such as quasi-stationary linear convective systems (a.k.a. Senjo-Kousuitai in Japan) that produce extreme rainfall (Kato 2020; Hirokawa et al. 2020). The other experiments were excluded from the scope of the present study because they were designed either as cloud-free-region seeding experiments to observe ice-nucleation processes (Jan. 7, 2026) or as experiments targeting quasi-stationary and stratiform cloud systems (Jan. 10 and 12, 2026). In addition, the present campaign represented the first cloud-seeding experiment conducted by our project. Therefore, the fourth experiment on 13 January was also the case in which coordination between decision support and aircraft operation was most successfully achieved, based on the feedback accumulated through the preceding experiments.

### 4.2 A Case of 13 January 2026

On 13 January 2026, the synoptic-scale environment was characterized by a wintertime pressure pattern, with a developing low-pressure system to the northeast of Japan and high pressure over the continent (Figure 5 a). This configuration favored the inflow of cold westerly air from the Sea of Japan, and was suitable for the formation of wintertime snow clouds around Toyama Bay. The radiosonde observation on the same day (Figure 5 b) showed a relatively moist layer from the lower troposphere to around the 700-hPa level, suggesting that cloud layers potentially suitable for seeding were likely to form at or below the operational altitude range of the aircraft. Above the lower moist layer, relatively dry air was present, except for possible mid-level moist layers. Ahead of the cold front, the target clouds were therefore considered to have developed primarily as shallow to moderately developed convective clouds in the lower to middle troposphere, rather than as deep cumulonimbus clouds. The temperature around the 700-hPa level was also sufficiently low to provide thermodynamic conditions suitable for glaciogenic seeding. Additionally, directional wind shear below approximately the 700-hPa level suggested that candidate clouds could move differently depending on their vertical extent. This provided another reason why real-time satellite and radar monitoring was necessary for target selection and aircraft guidance.

Note that, in this study, we distinguish between the “observation timeline”, which is based on the nominal observation times and does not include data latency, and the “actual operational timeline”, which accounts for the latency before the data became available to the ground team. In this subsection, we discuss the meteorological evolution using the observation timeline, focusing on the development and movement of the target clouds.

Figure 6 shows the time evolution of Himawari-9 visible imagery during the 13 January 2026 seeding in which cloud seeding was conducted for 3.5 min (11:15:20 – 11:18:50 JST). On this day, cloud formation was repeatedly observed ahead of the cold front (the blue dashed circles in Figure 6). This cloud region appeared quasi-stationary in terms of its geographical location, but the individual cloud elements within it were not stationary. Rather, they repeatedly formed, developed, and dissipated. This behavior is interpreted as the result of airflow crossing the Noto Peninsula that forced to updrafts in this region, thereby continuously generating shallow convective clouds. At the initial stage of the operation, this quasi-stationary cloud-generation region was considered a potential seeding target, and the aircraft therefore spent much of its flight time around this area (green lines in Figures 6 a-d). However, the ground team did not recommend seeding there because the candidate clouds did not satisfy the operational seedability criteria (Figure 7).

At 11:00:00 JST, a cumulus cloud initiated within the red-circled region and subsequently developed while moving northeastward under low-level southwesterly flow (Figures 6 d-g). The lifetime of this cell was approximately 20 min. It began to dissipate around 11:20:00 JST and had nearly disappeared by 11:25:00 JST (Figures 6 h and i, respectively). At the Himawari-9 observation time of 10:57:30 JST, the seedability assessment system did not yet identify this cumulus cloud, as indicated by the yellow dashed region (Figures 7 a-1 and b-1). By 11:00:00 JST, however, the cloud satisfied all three seedability criteria (Figures 7 a-2 and b-2). The subsequent observation at 11:02:30 JST showed a further decrease in B13 brightness temperature, indicating cloud-top cooling (Figure 7 a-3). Based on these features, the ground team identified this cloud as a suitable target for seeding. This information was communicated to the aircraft team at approximately 11:05:00 JST. Based on this guidance, the aircraft changed course (Figure 6 e), reached the target airspace at approximately 11:15:00 JST, and initiated seeding (Figure 6 g).

As described in Section 3.3, C-PAWR was also used for near-real-time confirmation of precipitation echoes around candidate clouds (Figure 8). Throughout the operation, C-PAWR did not show clear precipitation echoes associated with the quasi-stationary cloud indicated by the blue dashed ellipse in Figure 6. This was one of the reasons why the ground team did not recommend seeding this quasi-stationary cloud system. In contrast, although the target cloud system was outside the C-PAWR observation sector at 11:05:53 JST (Figure 8 a), it exhibited a clear radar echo at the 11:10:55 JST (Figure 8 b) after the aircraft had started moving toward the target cloud. This supported the interpretation that the target system was a precipitating convective cloud, with radar echoes located below the aircraft altitude, and provided additional evidence for the final target selection. The C-PAWR continued to monitor the target system until the start of seeding at 11:15:20 JST, thereby complementing the Himawari-9-based seedability assessment with local, near-real-time information on precipitation structure.

Figure 9 provides a three-dimensional overview of the spatial relationship among the C-PAWR reflectivity field, the aircraft track, the ground team location, and the voice guidance issued at approximately 11:05:00 JST. This visualization highlights the operational time lag between target-cloud identification by the ground team and aircraft arrival near the target cloud. This case demonstrates a clear example of how high-frequency Himawari-9 and C-PAWR observations enabled the aircraft to reach and seed a short-lived cumulus cloud with a lifetime of approximately 20 min. Figures 6 (h) and (i) show the subsequent dissipation of the seeded cloud. However, because the amount of seeding material released in this experiment was limited to 30 kg, we interpreted this dissipation as part of the natural evolution of the cloud rather than as a seeding-induced effect. More importantly, this case demonstrates that the proposed framework could guide the aircraft to a short-lived cloud and enable targeted seeding at an operationally favorable timing, immediately before the cloud began to dissipate naturally.

### 4.3 Importance of frequent remote-sensing observations

The present seeding operation was enabled by the availability of Himawari-9 observations at 2.5-min intervals. As described, even for geostationary satellite observations, an operational latency of approximately 5–10 min exists before the imagery becomes available to the ground team. To clarify how the seeding operation proceeded under this practical latency, Figure 10 summarizes the aircraft status and the corresponding seedability assessment system outputs along the "actual operational timeline" which includes the data latency. To distinguish these two timelines, "the actual operational timelines" are shown in italics in this section.

In the present case, the target cumulus cloud was not yet identifiable on the operational system at *11:02:30 JST* (corresponding to the Himawari-9's observation time of 10:57:30 JST). It was first detected on the system 2.5 min later (*11:05:00 JST*) and was immediately reported to the aircraft team as seedability guidance. After receiving this seedability report, the aircraft team required approximately 10 min to approach the target cloud and start the seeding operation. During the aircraft's approach to the target cloud, the C-PAWR echoes also confirmed that the target cloud was associated with a precipitating convective system. In this case, seeding started at *11:15:20 JST*, before the onset of dissipation of the target cumulus cloud. If the detection of the developing cumulus cloud had been delayed by approximately 10 min, for example until *11:15:00 JST*, the cloud would already have entered its dissipating stage by the time the aircraft reached the target (*11:25:00 JST*). In that case, the seeding operation would likely not have been successful because the target cloud had already dissipated.

This example demonstrates that the 2.5-min Himawari-9 observations were not merely useful for monitoring cloud evolution, but were essential for the timing of the operation. The 2.5-min rapid-scan observations are available over the Japan region and selected target areas, whereas Himawari-9 full-disk observations are generally provided at 10-min intervals. Therefore, the successful guidance in this case relied on the high-frequency regional observations provided by Himawari-9 satellite. The result highlights the importance of rapid-scan geostationary satellite observations for real-time target selection in aircraft-based cloud-seeding operations.

Even higher-frequency targeted observations may further improve the detection of rapidly evolving clouds. Himawari-8/9 have the capability to acquire Landmark Area imagery every 30 s (Bessho et al. 2016), and such 30-s observations have been used in recent studies of rapidly evolving tropical-cyclone inner-core structures (Horinouchi et al. 2023). However, their operational value for aircraft-based cloud seeding would depend not only on the imaging interval itself, but also on the total latency of data delivery, processing, interpretation, communication, and aircraft response.

## 5. Summary

This study aims to develop a real-time remote-sensing-guided decision-support framework for cloud-seeding operations, based on high-frequency remote-sensing observations. In particular, the seedability assessment system described in Section 3.2 proved useful for distinguishing candidate seedable clouds from the broader cloud field over Toyama Bay. The campaign demonstrated that high-frequency Himawari-9 and C-PAWR observations can be translated into actionable target selection and aircraft guidance for cloud-seeding operations under realistic field constraints.

Although Himawari-9 satellite provides 2.5-min observations over Japan, the imagery became available to the ground team with a delay of approximately 5–10 min. The aircraft also required time to reach candidate clouds, and the clouds visible from the aircraft were limited. Therefore, the operation required the ground team to anticipate which clouds would remain suitable and reachable for about 10–15 minutes later, rather than simply identify clouds that appeared suitable at the current time. These constraints limited the usefulness of conventional numerical weather predictions for minute-scale target selection. The field campaign therefore demonstrated the importance of the decision support framework that converts high-frequency satellite and radar observations into actionable guidance.

Weather intervention field campaigns require not only technical readiness but also procedural transparency and stakeholder communication. This is particularly important because intentional atmospheric intervention can raise public concerns regarding safety, environmental impacts, and possible unintended consequences, even when the experiment is conducted at a limited scale. The stakeholder communication process and public briefings conducted in this study therefore represent a practical example of a responsible pathway for advancing weather-intervention experiments in the present era.

Several limitations should be addressed for subsequent studies. First, this study did not establish a causal meteorological impact of cloud seeding. The amount of dry ice released was intentionally limited for safety reasons, and the expected seeding signal was likely much smaller than the natural variability of the snowfall system. This conservative interpretation is consistent with the long-recognized difficulty of attributing cloud-seeding signals against large natural variability in uncontrolled atmospheric systems (Garstang et al. 2005; Morrison et al. 2009; WMO 2025). Second, the sample size was small, and the campaign was conducted under location-specific wintertime conditions over Toyama Bay. The findings should therefore be interpreted as a field demonstration of operational feasibility rather than a statistical evaluation of seeding effectiveness.

Despite these limitations, the real-time decision-support framework developed in this study represents an important step toward operational weather-intervention research for mitigating the impacts of extreme rainfall. The framework demonstrates how high-frequency satellite and radar observations, human-in-the-loop target selection, aircraft guidance, and procedural transparency can be integrated into a single operational workflow. Such integration is essential for moving weather-intervention research from conceptual or numerical studies toward responsible field experimentation. Future studies should build on this framework by improving low-latency data delivery, aircraft position sharing, decision logging, AI-based cloud nowcasting, and seedability prediction. In parallel, field experiments with larger but carefully controlled seeding scales will be needed to evaluate whether overseeding can modify cloud microphysical processes in ways that suppress, redistribute, or mitigate intense precipitation. By progressively expanding the observational, operational, and experimental capabilities demonstrated here, future campaigns can contribute to establishing the scientific and technical basis for weather-intervention approaches aimed at reducing extreme rainfall hazards.

## Appendix A: Web-based seedability assessment tool

Here, we describe the web-based seedability assessment tool developed in this study using 2.5-min Himawari-9 data (see Section 3.2). During the Toyama Bay 2026 campaign, the system was automatically updated whenever new Himawari-9 data became available, allowing the ground team to monitor the latest cloud conditions in near real time.

Figure A1 shows the interface of the web-based seedability assessment tool. In addition to the Himawari-9-based display layers, the system provided operational overlays required for seeding decision making, including radar range rings, coastlines, grid labels, scale information, aircraft-related information, and the authorized airspace for dry-ice seeding. The grid labels were also used as a simple communication protocol between the ground and aircraft teams.

Instead of communicating target locations using latitude and longitude, the team referred to predefined grid numbers, such as A0 or B2, on the operational display. This approach was effective for sharing target locations reliably through satellite voice calls, where communication bandwidth and clarity were limited. In this way, the web-based system served not only as a visualization tool, but also as a shared operational reference for translating seedability information into concise aircraft guidance.

Figure A2 shows the three Himawari-9-based indices used for seedability assessment in this study: B13, B13−B15 and Tdiff in B13. These indices were used to monitor cloud-top temperature, cloud development or dissipation, and optically thick candidate clouds, respectively.

In the main text, we focus on two representative visualization modes of the assessment system: one that overlays B13 and B13-B15, and another that overlays all three Himawari-9-based seedability indices (e.g., Figures 4 a and b, respectively). These two modes were selected because they were most commonly used by the ground team during the campaign and were also useful for post-campaign analysis of cloud seeding.

## Abbreviations

AWS: Automatic weather station; C-PAWR: C-band phased-array weather radar; ELSI: Ethical,

legal, and social issues; HYVIS: Hydrometeor VideoSonde; JCSEPA: Japanese Cloud Seeding Experiments for Precipitation Augmentation; JMA: Japan Meteorological Agency; NWP: Numerical weather prediction; PPI: Plan-position indicator; RRI: Responsible research and innovation

## Declarations

### Funding

This research has been supported by JST Moonshot R&D (JPMJMS2389), and the IAAR Research Support Program and VL Program of Chiba University.

### Authors' contributions

SK and KY directed the project and the seeding experiments, and led the aircraft-based cloud-seeding operations. AM and KY directed the ground team and coordinated the operation and preparation of field observations. KS developed the Himawari-9-based seedability assessment system, and AT evaluated the decision-support framework using Himawari-9 data. YT participated in the aircraft-based cloud-seeding operations and contributed to improving the decision-support framework through coordination between the ground and aircraft teams. KK and YH optimized the experimental design in terms of the remote-sensing observation deployment and the aircraft-based seeding amount and altitude. TF and HK developed a new three-dimensional radar monitoring system and supported the experiment and subsequent analysis by integrating radar information with aircraft track data. Finally, KS, MO, and TT contributed to the overall logistics, implementation, and scientific evaluation of the experiment. KS promoted the RRI activities through coordination with the Toyama Prefectural Government and public briefings. MO enabled the experiments through the design of ground-based observation systems. TT helped ensure the scientific validity of the experiment through reviews of previous cloud-seeding experiments and interviews with relevant experts.

## Acknowledgements

The authors gratefully acknowledge Japan Radio Co., Ltd. (JRC) for providing and operating the C-band phased-array weather radar used in this study. We also thank Diamond Air Service Inc. (DAS) for their professional support in aircraft operations, including flight planning, operation, and safety management during the field campaign.

The authors are grateful to the project members who participated in the field campaign and supported its successful implementation. We also thank Dr. T. Hashino for conducting preliminary numerical weather prediction experiments to estimate the potential impacts of cloud seeding. We acknowledge the Moonshot Research and Development Program, particularly Drs. T. Nakazawa and K. Takatama, for conducting the experimental review process and providing guidance on the safe implementation of the campaign. We further thank the Toyama Prefectural Government for its support in communicating the campaign to local communities and stakeholders.

## Figures

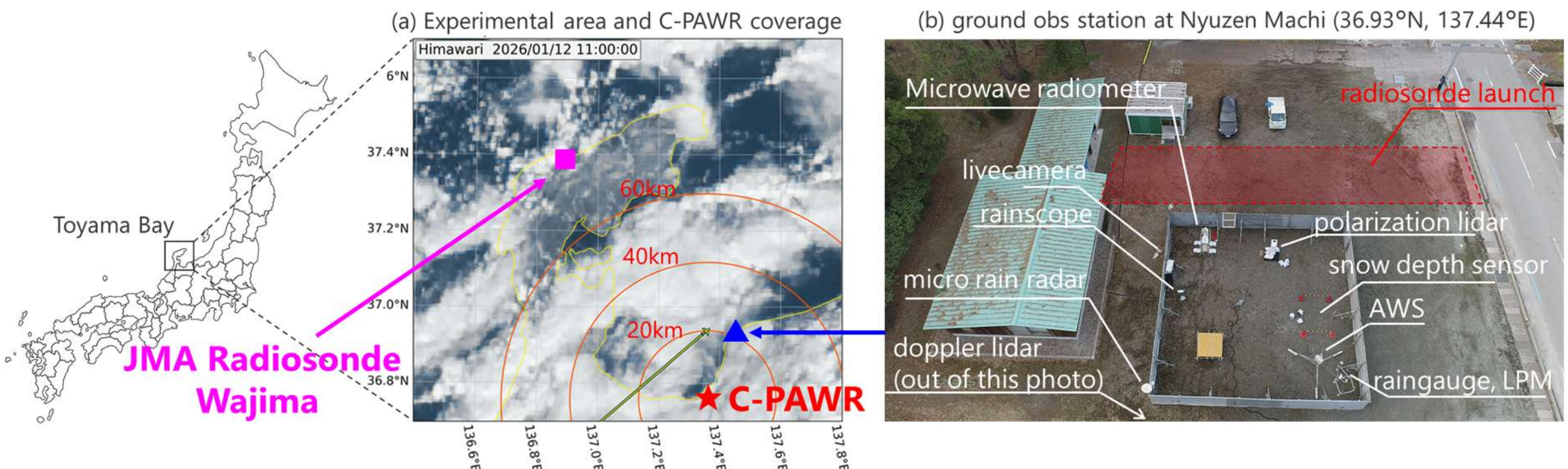

Figure 1. Field area and ground observation site of the Toyama Bay 2026 campaign. (a) Location of the field campaign around Toyama Bay. The left map shows the location of Toyama Bay in Japan, and the enlarged panel shows a representative Himawari-9 image over the experimental area. Red circles indicate the 20-, 40-, and 60-km range rings from the C-band Phased Array Weather Radar (C-PAWR) installed in Namerikawa. The blue triangle indicates the ground observation site in Nyuzen (36.93ºN, 137.44ºE).

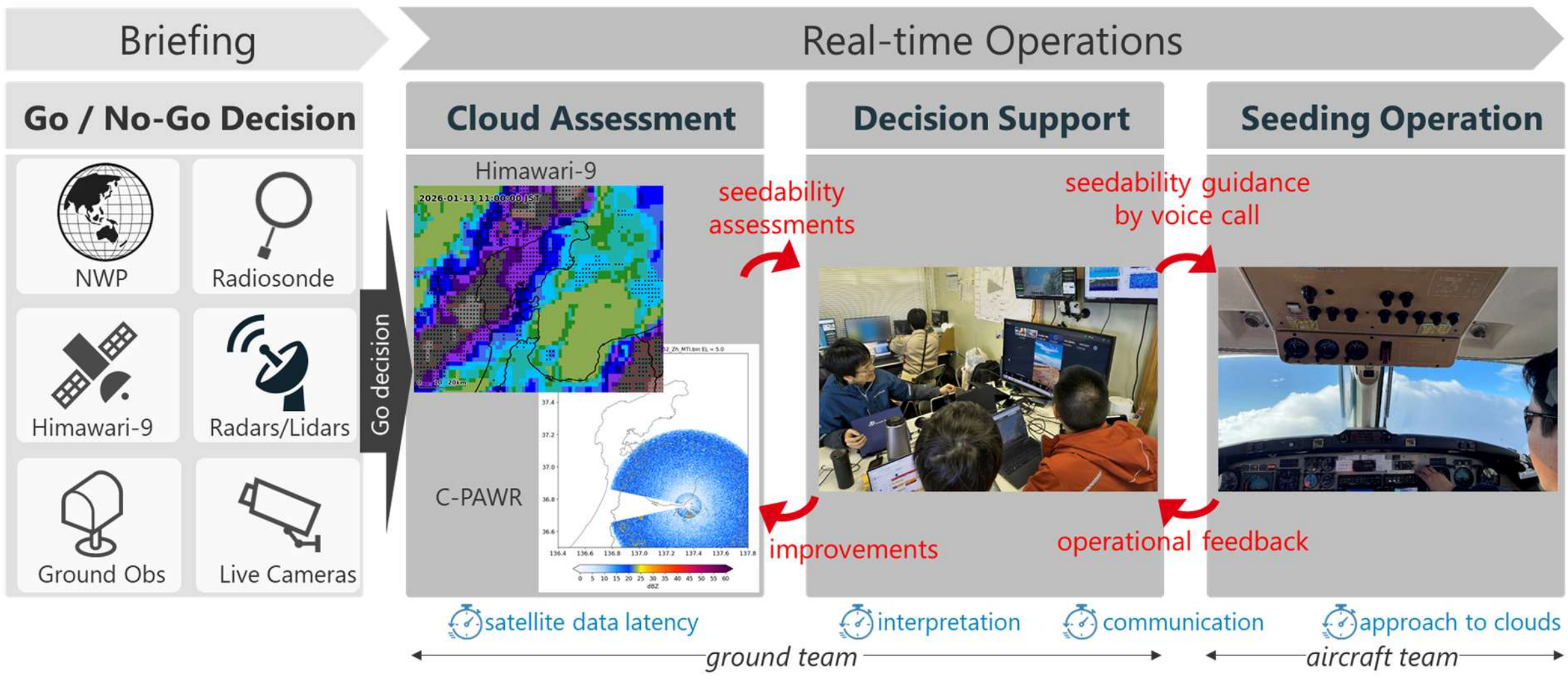

Figure 2: Real-time decision-support framework for cloud seeding operations. In practice, the decision-support process involves several operational time lags associated with satellite-data latency, ground-team interpretation, voice communication to the aircraft, and approach to the target cloud.

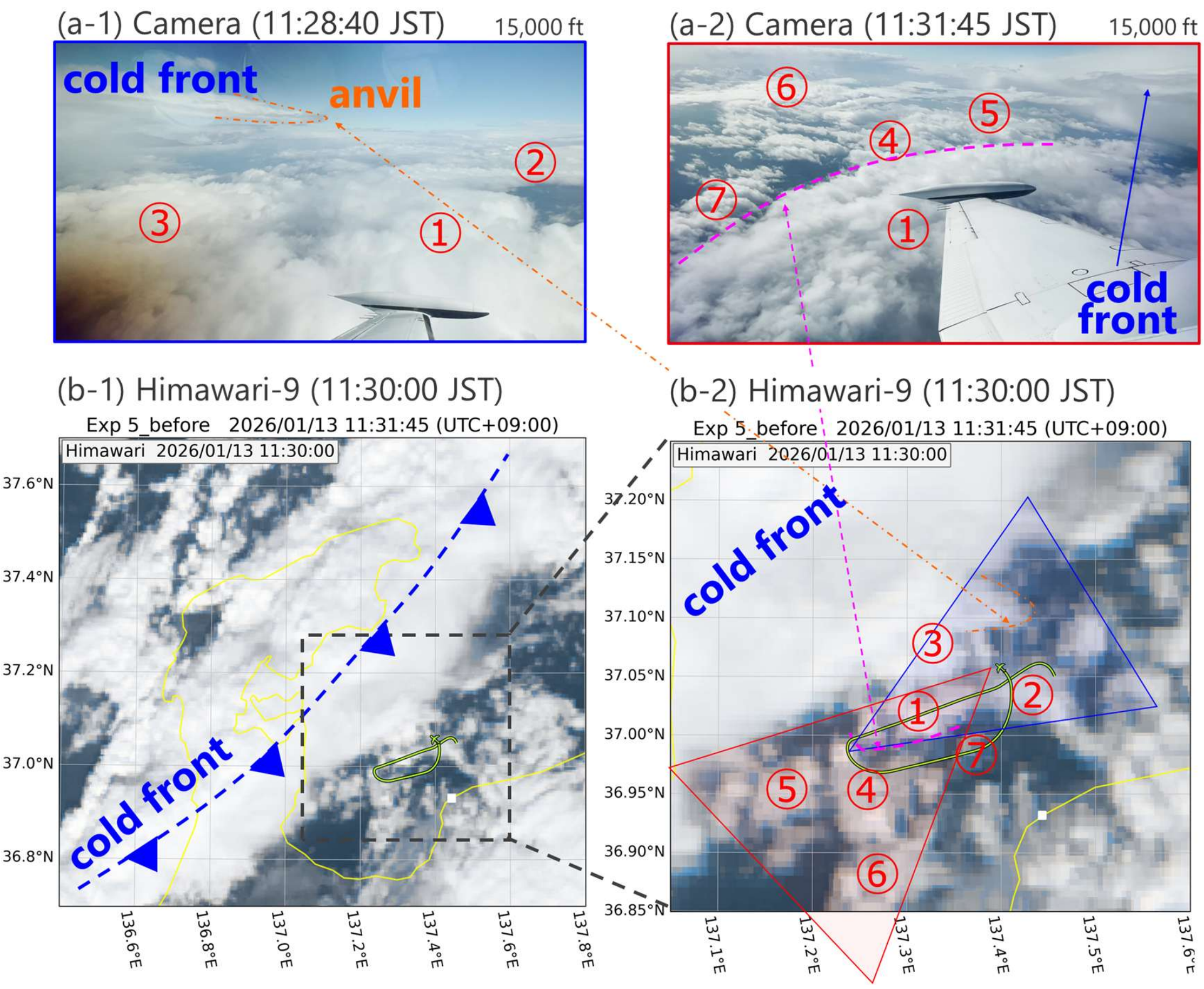

Figure 3. Correspondence between (a) aircraft visual observations and (b) Himawari-9 based visible image on 13 January 2026. Panels (a-1) and (a-2) show camera images taken from the aircraft at 11:28:40 and 11:31:45 JST, respectively at 4,572 m (15,000 ft). Panels (b-1) and (b-2) show the corresponding Himawari-9 imagery at 11:30 JST over the target region, with panel (b-2) providing an enlarged view. The numbered clouds in the aircraft images correspond to cloud features identified in the Himawari-9 imagery. The green line indicates the aircraft track, and the blue dashed line indicates the approximate position of the cold front. The red and blue triangles in panel (b-2) indicate the approximate fields of view from the aircraft corresponding to panels (a-1) and (a-2), respectively.

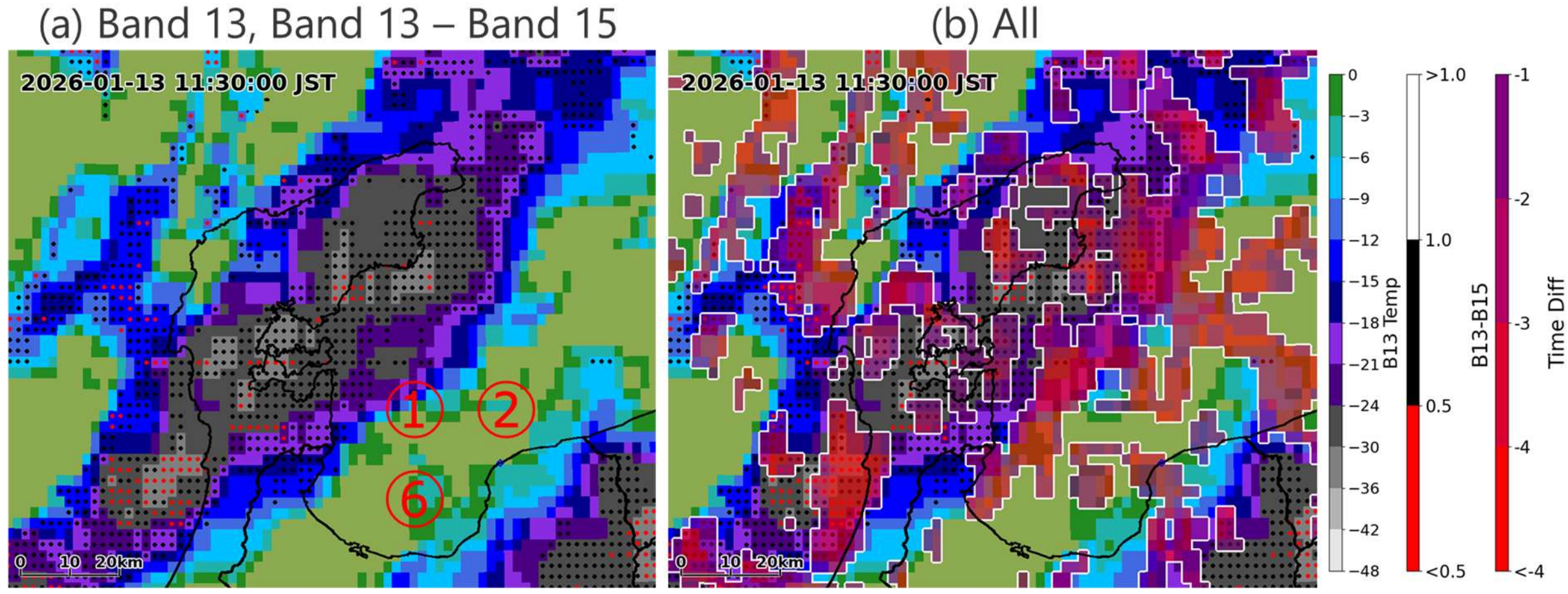

Figure 4. Himawari-9-based seedability assessments at 11:30 JST on 13 January 2026. (a) Composite display of B13 and B13−B15. (b) Composite display of all three Himawari-9-based seedability indices: B13, B13−B15, and Tdiff in B13. The numbered red circles indicate cloud features identified in the aircraft visual observations shown in Figure 3.

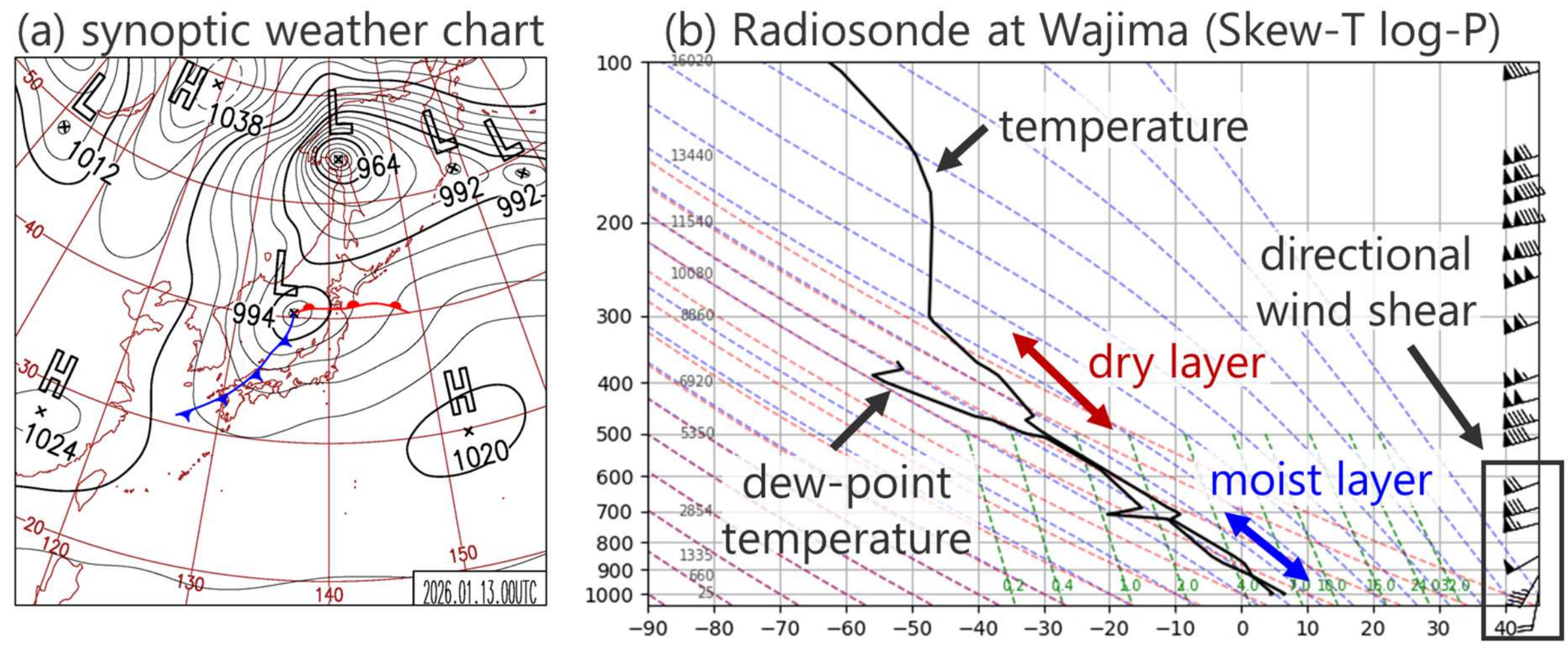

Figure 5 (a) Japan Meteorological Agency (JMA) synoptic weather chart and (b) radiosonde observation at Wajima (37.38ºN, 136.90ºE) for 0000 UTC from 13 January, 2026. The figures of weather chart and radiosonde are obtained from websites of JMA (https://www.data.jma.go.jp/) and The University of Wyoming (https://weather.uwyo.edu/upperair/sounding.shtml), respectively.

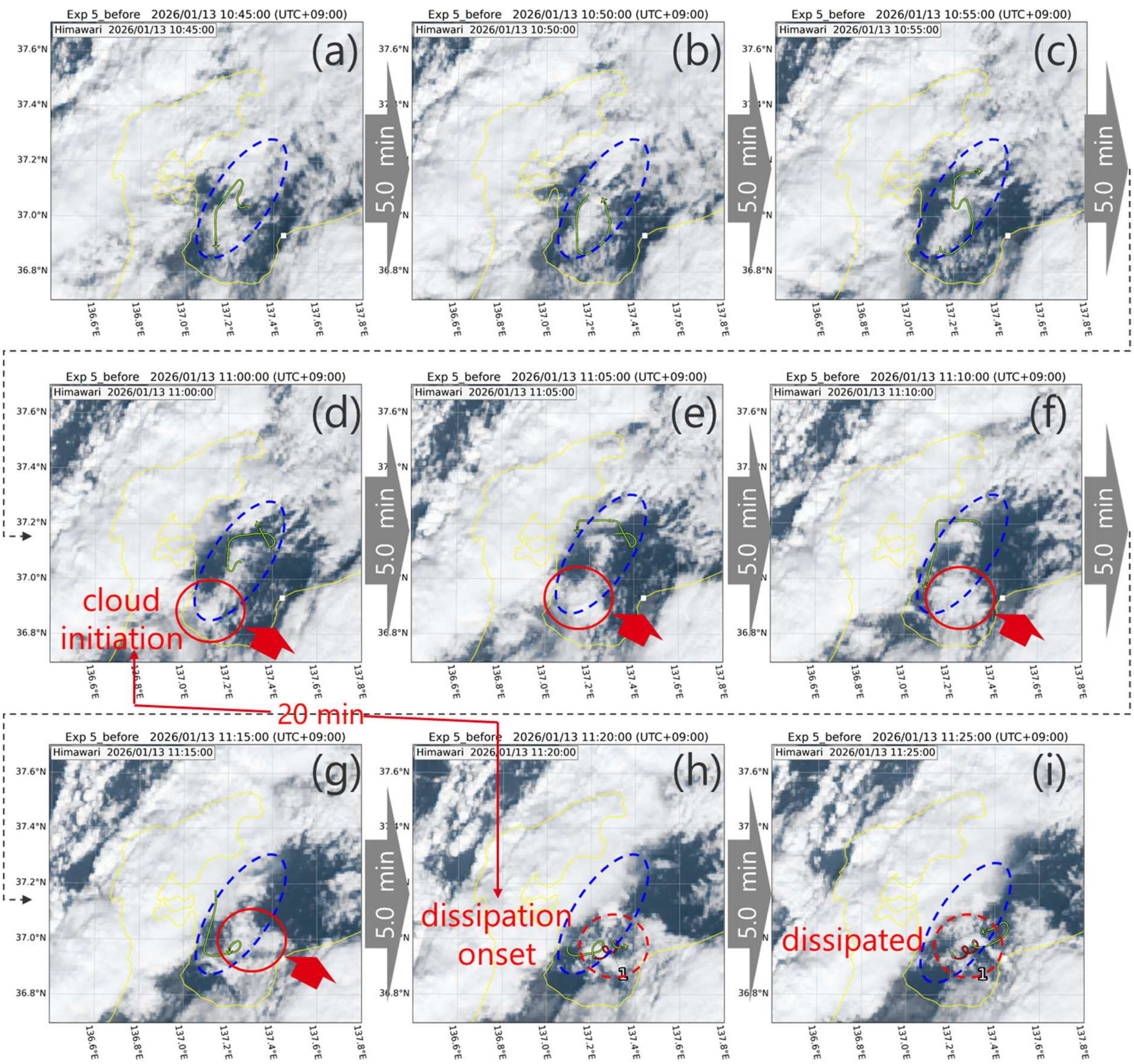

Figure 6. Time evolution of Himawari-9 visible imagery during the 13 January 2026 seeding operation. Panels (a)–(i) show Himawari-9 images from 10:45:00 to 11:25:00 JST (1:45:00 to 2:25:00 UTC) at 5-min intervals. The blue dashed ellipse indicates a quasi-stationary roll-like cloud band. The red circles indicate a developing cumulus cloud that approached the target region from the southwest and was selected as the seeding target by the ground team. The green lines show the last 10-min aircraft track.

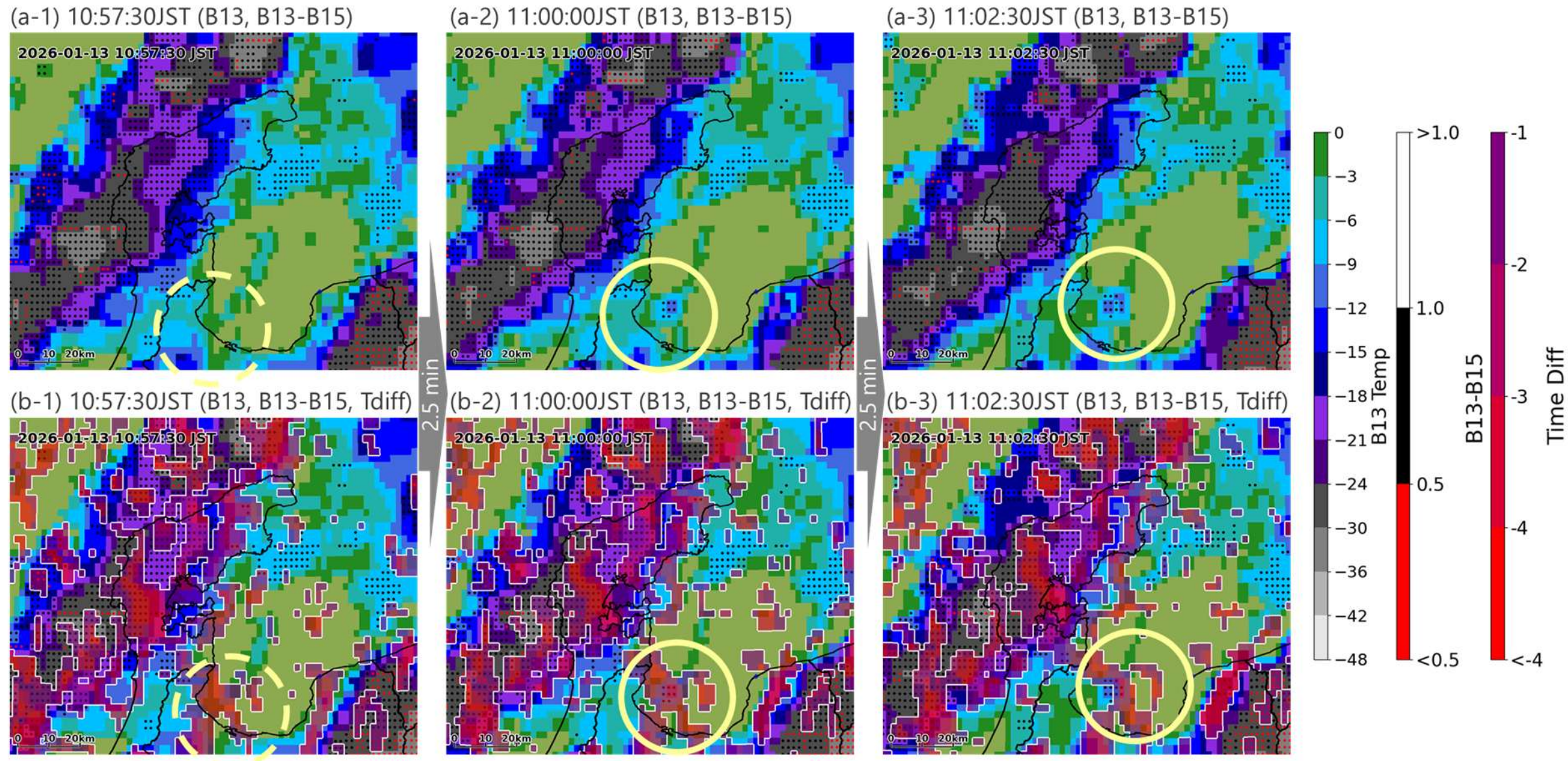

Figure 7. Time evolution of seedability assessment system from 10:57:30 to 11:02:30 JST (1:57:30 to 2:02:30 UTC) at 2.5-min intervals. (a) shows Himawari-9 B13 and B13−B15. (b) includes Tdiff in B13 in addition to (a).

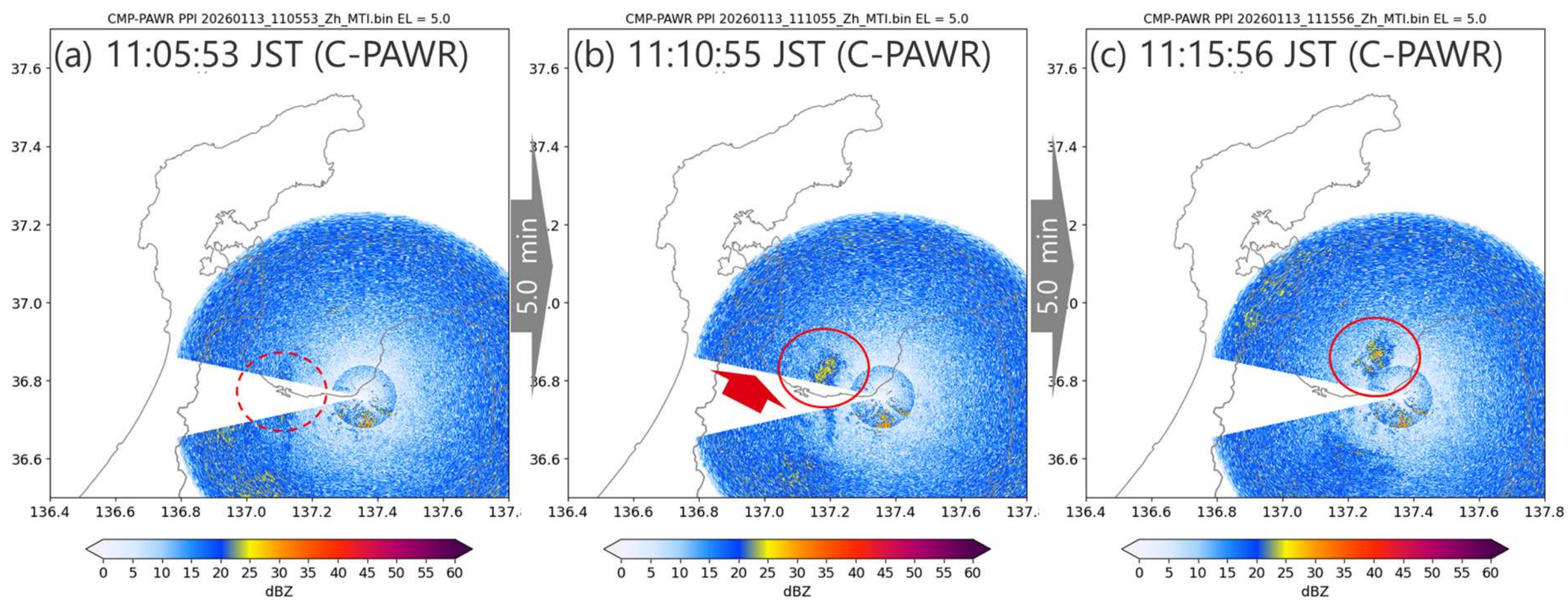

Figure 8. Time evolution of C-PAWR radar reflectivity during the 13 January 2026 seeding operation. Panels (a)–(c) show horizontal distributions of radar reflectivity at approximately 5-min intervals: (a) 11:05:53 JST (02:05:53 UTC), (b) 11:10:55 JST (02:10:55 UTC), and (c) 11:15:56 JST (02:15:56 UTC). The red circles indicate the precipitation echo associated with the developing cumulus cloud selected as the seeding target by the ground team. Panels (a)–(c) show plan-position indicator (PPI) scans of radar reflectivity at an elevation angle of 5.0°. The white region to the west of the radar indicates an area not observed because of beam blockage by surrounding obstacles.

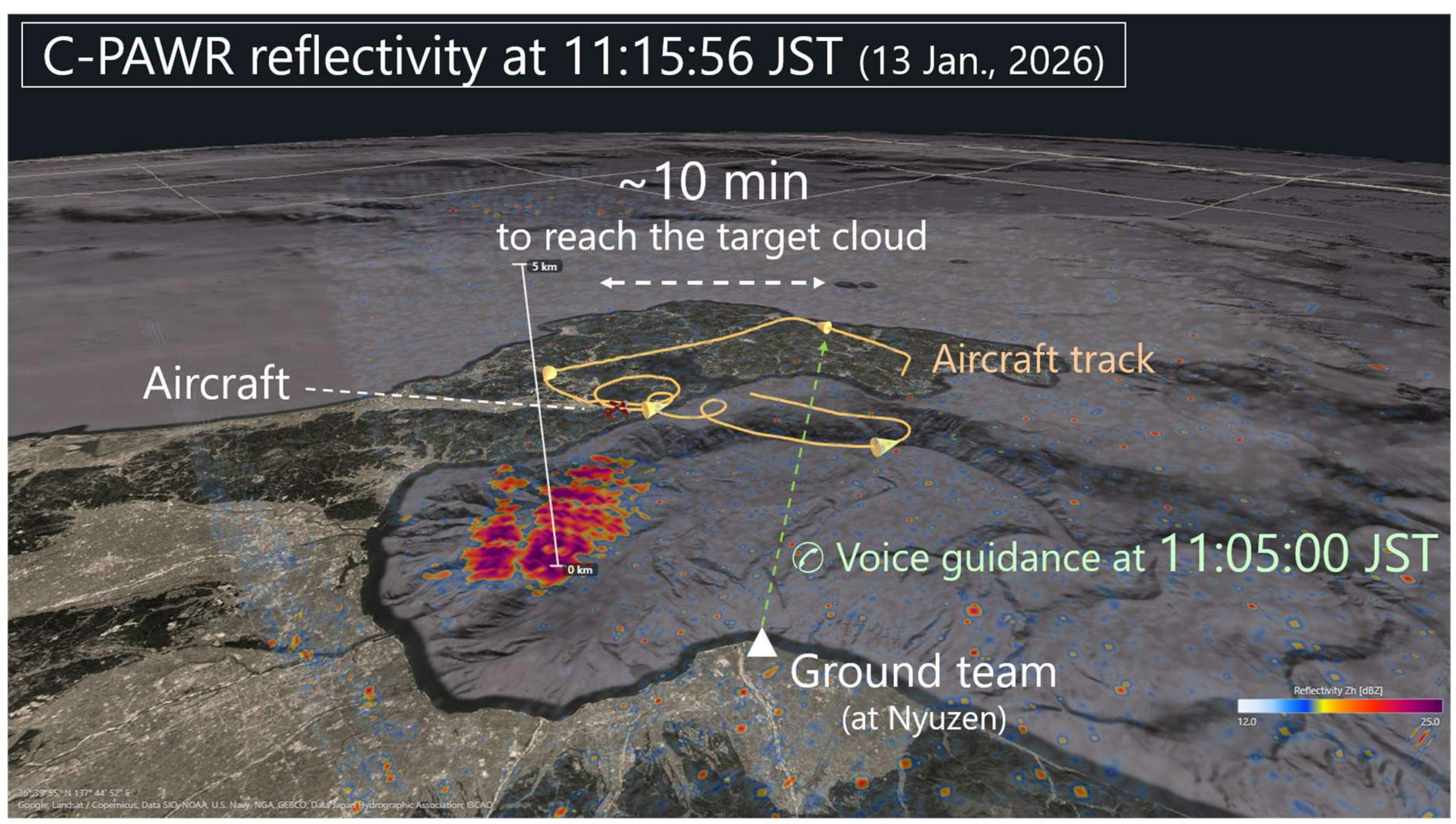

Figure 9. Three-dimensional overview of the 13 January 2026 cloud-seeding operation over Toyama Bay. The C-PAWR reflectivity field at 11:15:56 JST is shown together with the aircraft track, the location of the ground team at Nyuzen, and the voice guidance issued at approximately 11:05:00 JST.

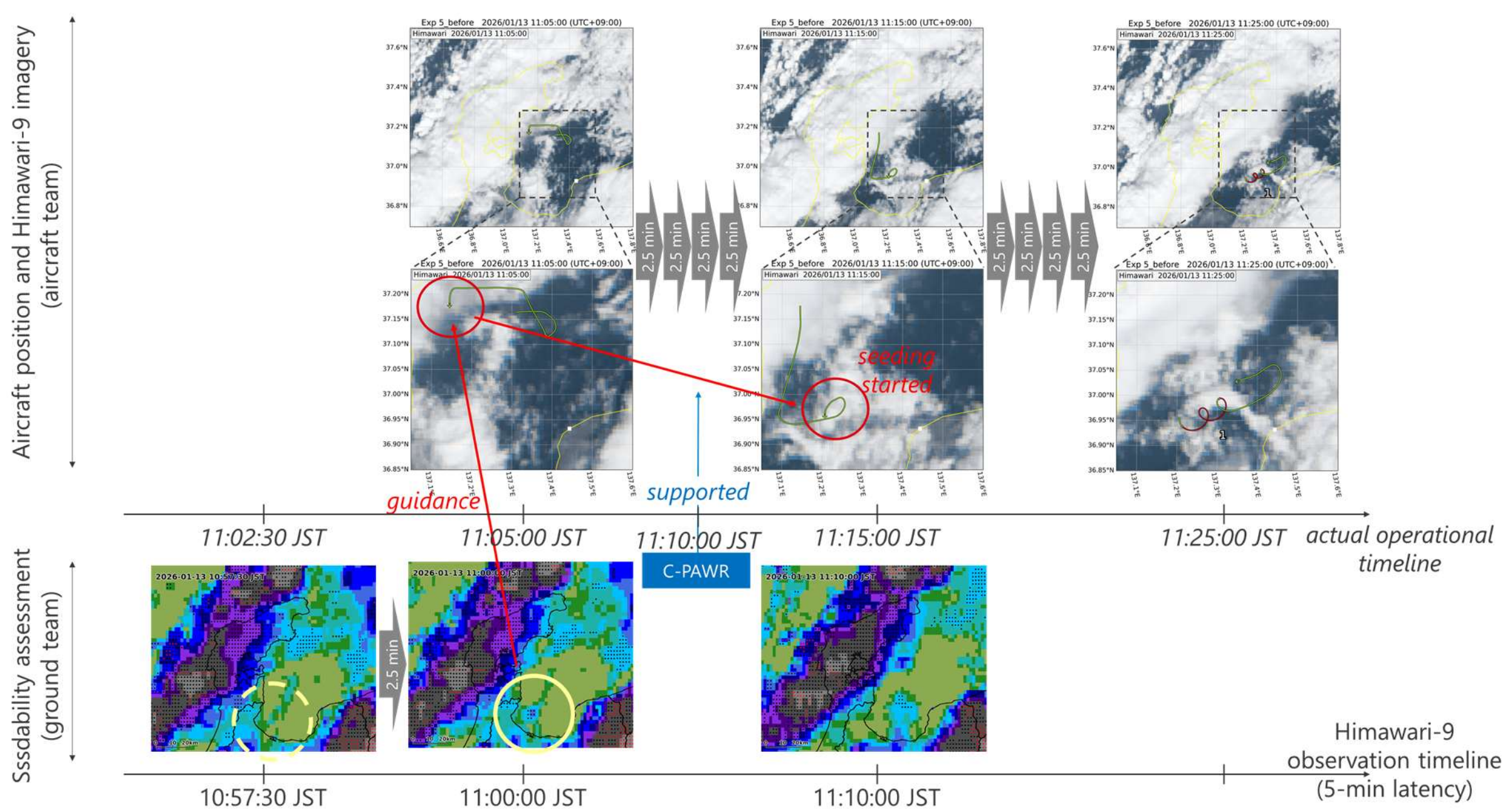

Figure 10. Real-time target selection and aircraft guidance based on Himawari-9 observations during the 13 January 2026 seeding operation along the actual operational timeline. The upper panels show successive Himawari-9 images around the target region, and the lower enlarged panels highlight the cloud evolution. The red annotations indicate the guidance process and the timing at which seeding started, while the green lines show the aircraft track. The lower panels show the operational seedability display used by the ground team.

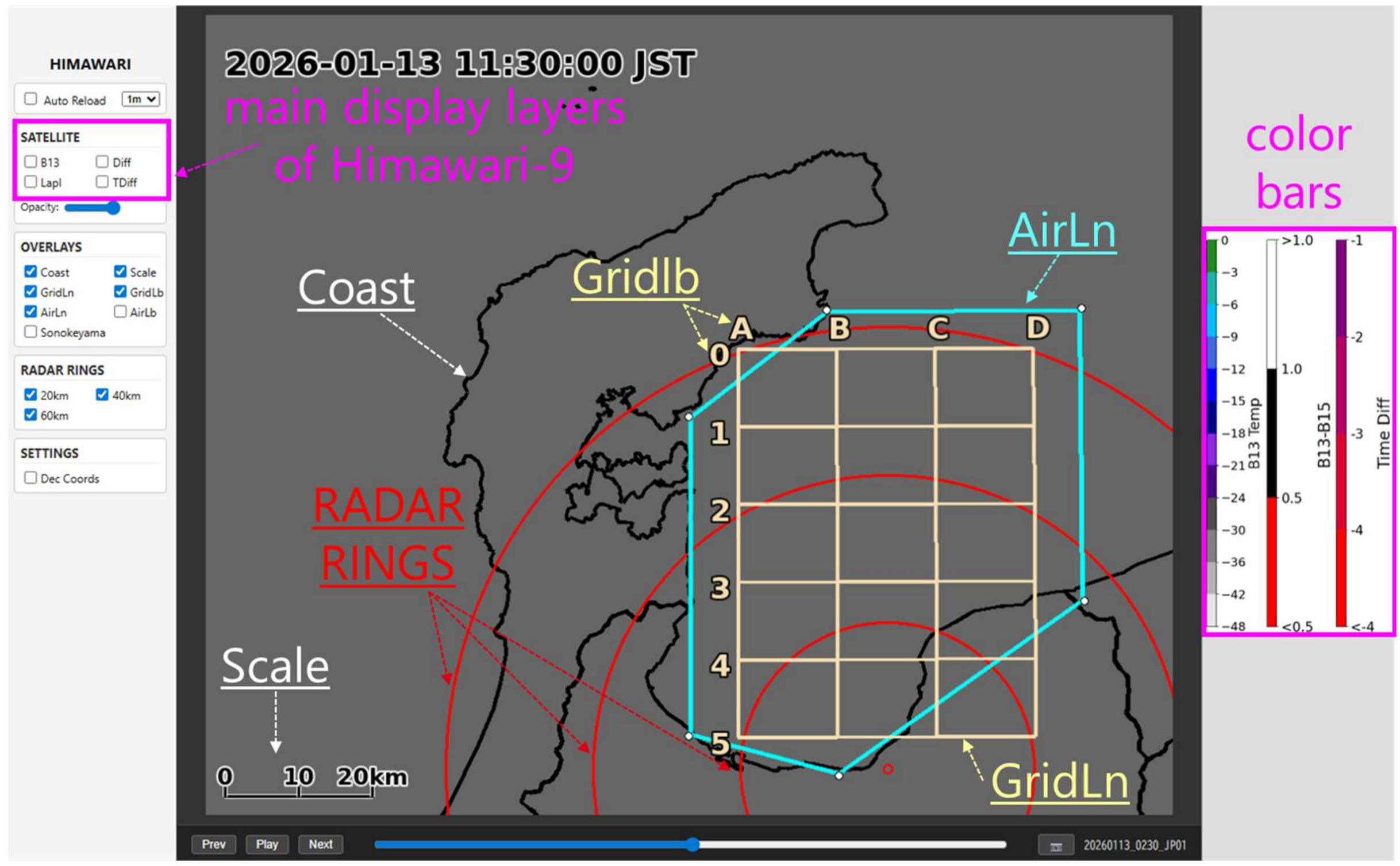

Figure A1: An overview of the web-based seedability assessment tool used during the Toyama Bay 2026 campaign on 11:30:00 JST of 13 January, 2026. The display combined Himawari-9-based seedability indices with operational overlays, including coastlines, grid labels, aircraft-related information, radar range rings, and a scale bar. The screenshot shows the default display without Himawari-9 data. The cyan polygon indicates the airspace in which dry-ice seeding was permitted. The selectable Himawari-9 layers include B13, B13−B15 and temporal difference in B13.

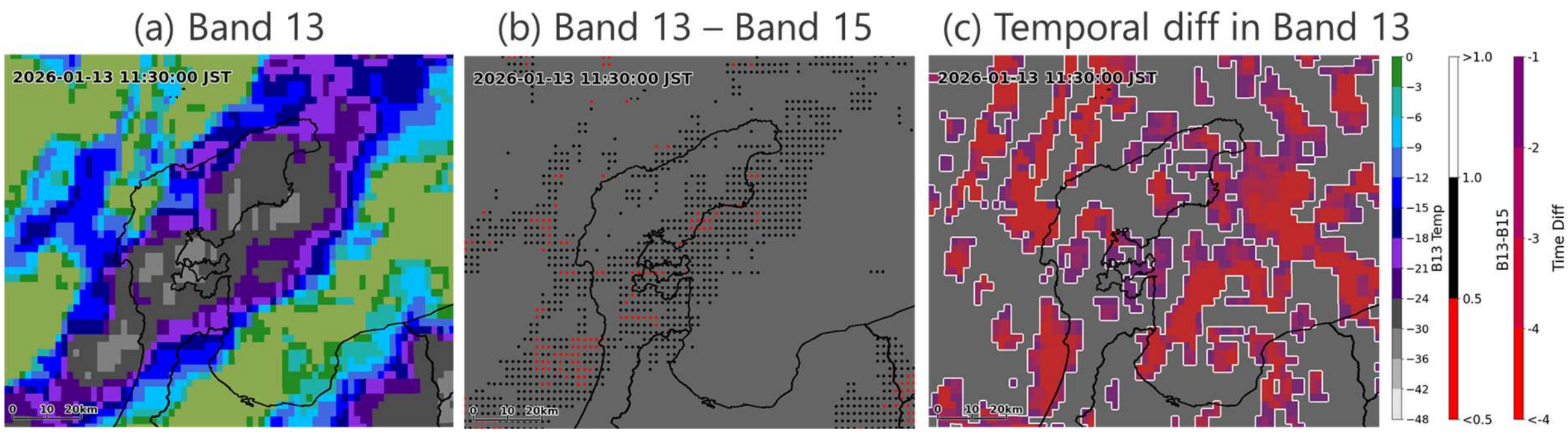

Figure A2. Himawari-9-based seedability indicators at 11:30 JST on 13 January 2026. The panels show (a) B13, (b) B13−B15 and (c) temporal difference in B13.

## Tables

Table 1. Daily summary of the Toyama Bay 2026 campaign. Seeding operations were conducted on 7, 10, 12, and 13 January under different cloud conditions. Non-operational days were determined by the Go/No-Go criteria, including wind direction, weather conditions, and lightning risk.

| Date | seeding operations | Reason for No-Go Decision | Flight time |
|---|---|---|---|
| Jan 6, 2026 | | unsuitable wind direction | |
| Jan 7, 2026 | cloud-free region seeding | | 3 h 35 min |
| Jan 8, 2026 | | unsuitable wind direction | |
| Jan 9, 2026 | | unsuitable seeding condition | |
| Jan 10, 2026 | stationary-cloud seeding | | 4 h 5 min |
| Jan 11, 2026 | | high lightning risk | |
| Jan 12, 2026 | stratiform-cloud seeding | | 2 h 35 min |
| Jan 13, 2026 | convective-cloud seeding | | 3 h 45 min |
| Jan 14, 2026 | | unsuitable wind direction | |